\documentclass[12pt,preprint]{aastex}

\newcommand{\teff}{$T_{\mathrm{eff}}$}
\newcommand{\teffs}{$T_{\mathrm{eff}} \;$}

\begin{document}

\title{HIGH-RESOLUTION SPECTROSCOPY OF THE PLANETARY HOST HD 13189: 
HIGHLY-EVOLVED AND METAL-POOR\altaffilmark{1}}

\altaffiltext{1}{Based on observations obtained at the 2-m Alfred Jensch 
telescope at the Th{\"u}ringer Landessternwarte Tautenburg, Tautenburg, 
Germany.}

\author{Simon C. Schuler\altaffilmark{2}, James H. Kim \altaffilmark{2,3}, 
Michael C. Tinker, Jr.\altaffilmark{2}, Jeremy R. King\altaffilmark{2}, 
Artie P. Hatzes\altaffilmark{4}, AND Eike W. Guenther\altaffilmark{4}}
\affil{
  \altaffiltext{2}{Department of Physics and Astronomy, Clemson University, 118
  Kinard Laboratory, Clemson, SC, 29634; sschule@ces.clemson.edu,
  jking2@clemson.edu, mtinker@ces.clemson.edu}
  \altaffiltext{3}{Department of Astronomy, Boston University, 725 Commonwealth 
  Ave., Boston, MA, 02215; jimk818@bu.edu}
  \altaffiltext{4}{Th{\"u}ringer Landessternwarte Tautenburg, Sternwarte 5, 
  D-07778 Tautenburg, Germany; artie@tls-tautenburg.de, 
  guenther@tls-tautenburg.de}}
 
\begin{abstract}
We report on the abundances of 13 elements in the planetary host HD 13189, a 
massive giant star.  Abundances are found to be sub-solar, with ${\rm [Fe/H]} = 
-0.58 \pm 0.04$; HD 13189 is one of the most metal-poor planetary hosts yet 
discovered.  Abundance ratios relative to Fe show no peculiarities with respect to 
random field stars.  A census of metallicities of the seven currently known 
planet-harboring giants results in a distribution that is more metal-poor than the 
well-known metal-rich distribution of main sequence (MS) planetary hosts.  This 
finding is discussed in terms of accretion of H-depleted material, one of the 
possible mechanisms responsible for the high-metallicity distribution of MS stars 
with planets.  We estimate the mass of the HD 13189 progenitor to be $3.5 \; 
M_{\sun}$ but cannot constrain this value to better than 2-6 $M_{\sun}$.  A stellar 
mass of $3.5 \; M_{\sun}$ implies a planetary mass of $m \sin i = 14.0 \pm 0.8 \; 
M_{J}$, placing the companion at the planet/brown dwarf boundary.  Given its 
physical characteristics, the HD 13189 system is potentially unique among planetary 
systems, and its continued investigation should provide invaluable data to 
extrasolar planetary research.
\end{abstract}


\section{INTRODUCTION}
Ascertaining the physical properties of planetary host stars, the chemical
abundances thereof in particular, is a critical component of understanding the 
formation and evolution of extrasolar planetary systems.  Of the approximately 
131 stars presently known to harbor planets, Fe abundances have been derived for 
no fewer than 117, with the abundances of numerous other elements having been 
determined for many.  The results of these abundance analyses have led to the now 
well-known discovery that stars with planets tend to be metal-rich compared to 
random field stars (e.g., Fischer \& Valenti 2005; Santos et al. 2005).  Two 
hypotheses have emerged as possible explanations of the planet-metallicity 
correlation: accretion of H-depleted, rocky material by planetary hosts and 
preferential planet formation in high-metallicity, proto-planetary disks.  Both 
of these propositions are discussed extensively in the pertinent literature 
(e.g., Gonzalez 1997; Ida \& Lin 2004; Fischer \& Valenti 2005).

An interesting consequence of the radial velocity (RV) method used currently to 
detect extrasolar planets is the limited range of stellar spectral types of stars 
chosen as targets.  Planet searches generally focus on older, main sequence 
(MS) late-F, G, and K dwarfs, because it is these stars that are bright enough to
obtain high signal-to-noise, high-resolution spectra, have an ample number of usable
spectral lines for the RV analyses, and have rotation rates and activity levels 
that facilitate the detection of planets.  Furthermore, there is potentially great 
sociological significance in finding planets around stars like our Sun.  While 
unique challenges exist for finding planets around younger stars 
and stars of earlier and later spectral types (e.g., Setiawan et al. 2003; 
Paulson, Cochran, \& Hatzes 2004), the paucity of known planets in orbits around 
these ``other'' stars prohibits a full elucidation of extrasolar planet formation 
and evolution.

Recognizing the importance of filling the gaps of the currently known planetary 
host sample, a handful of groups have initiated dedicated surveys of low mass stars 
\citep{2003AJ....126.3099E} and evolved G- and K-giants, the progenitors of which 
were massive early-type MS dwarfs \citep{2003toed.conf..441H}.  Most recently, the 
Tautenburg Observatory Planet Search (TOPS) program, using the Alfred-Jensch 2-m 
telescope at the Th{\"u}ringer Landessternwarte Tautenburg (TLS) with follow-up 
observations with the 2.7-m Harlan J. Smith and Hobby Eberly (HET) telescopes at 
The McDonald Observatory, has announced the discovery of a giant planet orbiting 
the K2 II giant \object{HD 13189} (Hatzes et al. 2005; Paper I).  Based on an 
estimated luminosity of 3.6 $L_{\sun}$ and the evolutionary tracks of 
\citet{1996A&AS..117..113G}, Paper I estimated a mass range of 2-7 $M_{\sun}$ for 
the MS progenitor of HD 13189.  Thus, HD 13189 may be the most massive star known 
to harbor a planet.  Here, we present the results of our abundance analysis of this 
potentially unique planetary host.

\section{OBSERVATIONS AND DATA ANALYSIS} 
Our abundance analysis makes use of two template spectra (spectra taken without an 
iodine cell) obtained as part of the TOPS program on 2002 December 19 using the 
Alfred-Jensch 2-m telescope and the high-resolution coud{\'e} echelle spectrometer 
at TLS, located in Tautenburg, Germany.  The spectrometer consists of an echelle 
grating with $31.6 \; \mathrm{g} \; \mathrm{mm}^{-1}$ and a cross-dispersing grism; 
the VISUAL grism has been used, along with a single 2048 x 2048 CCD detector with 
$15 \; \micron$ pixels, providing a wavelength coverage of 4660-7410 {\AA}.  Each 
spectrum has a resolution of $R = 67,000$ and a typical signal-to-noise (S/N) ratio 
of $\sim 115$.  A standard data reduction process, which includes bias subtraction, 
flatfielding, scattered light removal, extraction, and wavelength calibration, 
using the usual routines within the {\sf IRAF} facility has been applied to the 
spectra.

The two individual HD 13189 spectra were co-added into a single, higher S/N 
ratio ($\mathrm{S/N} \approx 160$) spectrum, and equivalent widths (EWs) were 
measured by fitting lines with Gaussian or Voigt profiles using the one-dimensional 
spectrum analysis package {\sf SPECTRE} \citep{1987BAAS...19.1129F}.  Spectral line 
selection was aided by the collection of \citet{1990A&AS...82..179T}; only lines 
labeled as ``case a''- lines with estimated internal uncertainties of $\leq 0.05 \; 
\mathrm{dex}$ in oscillator strengths ($\log gf$) and presumably free from blends 
in the solar spectrum- were considered.  The list was culled further by examining 
each line in the spectrum of HD 13189 and eliminating those that were deemed to be 
blended or otherwise unmeasurable.  Our final line list, along with the measured 
EWs for both HD 13189 and the Sun, is presented in Table 1.  Solar EWs have been 
measured in a high-quality ($R = 60,000$ and $\mathrm{S/N} = 1,000$) spectrum of 
the daytime sky obtained with the 2.7-m Harlan J. Smith telescope at the McDonald 
Observatory \citep{trip}.


Stellar parameters have been determined spectroscopically via an iterative 
process.  The \teffs and $\xi$ are determined by adjusting their values until there 
are no correlations between line-by-line [Fe/H]\footnotemark[5] abundances (as 
derived from \ion{Fe}{1} lines) and excitation potential ($\chi$) and reduced EW 
($\log \mathrm{EW}/\lambda$), respectively.  This method for determining \teffs and 
$\xi$ is used often, but care must be taken to ensure that there is no ab 
initio correlation between excitation potential and reduced EW, which can lead 
to degenerate solutions.  This is the case with our \ion{Fe}{1} list, which 
consists of 86 lines.  The list was trimmed until the correlation between 
excitation potential and reduced EW was eliminated; this left 47 \ion{Fe}{1} 
lines with a satisfactory range of excitation potentials ($0.91 \leq \chi \leq 
4.99 \; \mathrm{eV}$).  Uncertainties in \teffs and $\xi$ ($1\sigma$) are 
determined by adjusting the parameters until the corresponding correlation 
coefficient has a $1\sigma$ significance.  Finally, $\log g$ is fixed by 
forcing Fe abundances derived from \ion{Fe}{1} and \ion{Fe}{2} lines into
agreement; the uncertainty in $\log g$ is based on the uncertainty in the 
difference of the \ion{Fe}{1} and \ion{Fe}{2} abundances.  The final stellar 
parameters and uncertainties are given in Table 2.  We note our derived 
$\log g$ value supports the finding of Paper I that HD 13189 is a highly 
evolved giant.


\footnotetext[5]{The bracket notation is used to denote abundances relative to 
solar values, e.g., $\mathrm{[Fe/H]} = \log \{N(Fe)/N(H)\}_{\star} - \log 
\{N(Fe)/N(H)\}_{\sun}$, where $\log N(H) = 12.0$.}

Abundances have been derived using the LTE stellar line analysis package {\sf MOOG} 
(Sneden 1973; Sneden 2004, private communication).  The {\sf abfind} driver was 
used to force-fit abundances to the line-by-line EWs for all elements except O.  We 
have used the {\sf blends} driver to derive the abundance of O from the 
$\lambda 6300$ [\ion{O}{1}] line, taking care to account for the non-negligible 
contribution to the feature by a \ion{Ni}{1} blend (e.g., Allende Prieto et al. 
2001).  The specifics of our O analysis follows that of \citet{forb}, which should 
be consulted for details.  Model atmospheres with the convective overshoot 
approximation for HD 13189 and the Sun have been interpolated from the ATLAS9 grids 
of Kurucz; the solar parameters $T_{\mathrm{eff}} = 5777 \; \mathrm{K}$, 
$\log g = 4.44$, and $\xi = 1.25$ were adopted.  Atomic parameters for all 
transitions are from \citet{1990A&AS...82..179T}, except for those of Mg, which are 
from the VALD database (Piskunov et al. 1995; Kupka et al. 1999; Ryabchikova et al. 
1999), and that of $\lambda 6300$ [\ion{O}{1}], which is from 
\citet{2001ApJ...556L..63A}. 

Final absolute abundances for HD 13189 and the Sun are given in Table 1, and 
the final stellar parameters, relative abundances, and uncertainties for HD 
13189 are presented in Table 2.  The \ion{Fe}{1} abundances are derived with 
the complete list of 86 lines.  Relative abundances are the mean of a 
line-by-line comparison with solar values, thus limiting the impact of 
possibly inaccurate $gf$-values.  Final abundance uncertainties are the quadratic
sum of the uncertainties related to the adopted stellar parameters- based on 
sensitivities to arbitrary changes in \teff, $\log g$, and $\xi$- (Table
3)- and uncertainties in the mean abundances.  For Mn and O 
abundances, which are based on a single feature, the final uncertainties 
incorporate uncertainties in the solar abundances.  Sensitivities to the 
adopted C (important for molecular equilibrium calculations, primarily related 
to CO) and Ni abundances are also included in the final O abundance 
uncertainty.  The final abundance uncertainties are on the order of 0.05-0.10 dex,
but larger uncertainties due to systematic errors, e.g., related to relative
differences in the atmospheric models, cannot be ruled out.


\section{DISCUSSION}
The derived Fe abundance, [Fe/H] $= -0.58 \pm 0.04$, places HD 13189 in the 
lower envelope of the planetary host metallicity distribution and makes HD 
13189 one of the most metal-poor planetary hosts yet discovered.  Furthermore,
with a semi-major axis in the range 1.5-2.2 AU (Paper I), the planetary
companion of HD 13189 is the only one known to have a semi-major axis $> 1.0$ 
AU and orbit a star with a metallicity $< -0.40$, a combination that is 
potentially important to understanding giant planet formation 
\citep{2005ApJ...630.1107R}.  Abundances of other elements considered here are 
sub-solar, and the [m/Fe] ratios are unspectacular when compared to the general 
field population (e.g., Edvardsson et al. 1993; Fulbright 2002).  The 
planet-metallicity correlation is one of the most scrutinized results to emerge 
from extrasolar planet studies.  \citet{2005ApJ...622.1102F} offer the most 
thorough investigation of the correlation to date; their conclusion, based on a 
uniform analysis of 850 stars with and without planets, is that gas giant planets 
preferentially form in high-metallicity environments.  Various groups have also 
argued for this ``primordial'' scenario (e.g., Pinsonneault, DePoy, \& Coffee 2001; 
Santos 2003), and there are two main pieces of evidence given in support.  First, 
there is no correlation between convection zone (CZ) depth and metallicity of 
planetary hosts, an effect that is expected if accretion of H-depleted rocky 
material has occurred.  Second, the metallicity distribution of planet-harboring 
subgiant stars does not differ from that of dwarfs; the deepening CZ of a subgiant 
should dilute any accretion signatures that might have occurred.
  
Despite the substantial evidence pointing to the primordial scenario for planetary
formation, data in support of the accretion of H-depleted rocky material
hypothesis also exist, albeit they are more sparse.  \citet{1997MNRAS.285..403G} 
was the first to report on the super-solar metallicities of planetary hosts and 
to suggest the correlation may be a result of accretion.  Shortly thereafter, 
\citet{2001Natur.411..163I} detected $^6{\rm Li}$ in the metal-rich planetary 
host HD 82943; $^6{\rm Li}$ is a volatile species that should be destroyed beyond 
the level of observability in the atmospheres of solar-type stars during 
pre-MS evolution (e.g., Proffitt \& Michaud 1989).  Israelian et al. interpreted 
the presence of $^6{\rm Li}$- which should be preserved in giant planets- in the 
HD 82943 as possible evidence of accretion; the possibility of heightened $^6$Li 
abundances in the atmospheres of stars that have engulfed rocky material was later 
confirmed by \citet{2002A&A...386.1039M}.  The $^6{\rm Li}$ detection was 
subsequently challenged by \citet{2002MNRAS.335.1005R}, who questioned the atomic 
data used by Israelian et al.  In response, \citep{2003A&A...405..753I} reanalyzed 
$^6{\rm Li}$ in HD 82943 using updated atomic data and confirmed their original 
detection.  Further evidence for the accretion hypothesis comes from the study of 
Smith, Cunha, \& Lazzaro (2001), who found a trend of increasing abundance for 
elements with increasing condensation temperatures ($T_{\rm c}$) for a small 
subsample of planetary hosts.  If accretion of rocky material has occurred, one 
might expect the material to be rich in high-$T_{\rm c}$, refractory elements 
compared to low-$T_{\rm c}$, volatile elements \citep{1997MNRAS.285..403G}.  
However, similar $T_{\rm c}$-dependent abundance trends in planetary hosts have not 
been found by others (e.g., Sadakane et al. 2002).

Additional support for the accretion hypothesis may be emerging from the discovery 
and analyses of giant stars with planetary companions.  As of this writing, planets 
are known to orbit two G-type and five K-type giants (including HD 13189), and 
their metallicity distribution is not similar to that of planet-hosting dwarfs.  
In Figure 1 we plot relative Fe abundances versus \teffs for the seven giants (red 
circles) and for planet hosting dwarfs from Fischer \& Valenti (2005; black 
circles), from Santos, Israelian, \& Mayor (2004; black squares), excluding 
duplicates with Fischer \& Valenti, and from Santos et al. (2005; black triangles), 
again excluding duplicates.  The horizontal line represents the mean [Fe/H] 
abundance of the combined, 126 dwarf sample.  The giant data for HD 47536, 
HD 59686, HD 137759, and HD 219449 are from \citet{2005PASJ...57..127S}, and data 
for HD 11977 and HD 104985 are from \citet{2005A&A...437L..31S} and 
\citet{2003ApJ...597L.157S}, respectively.  In the combined dwarf sample, 
$41.3\%$ have derived [Fe/H] abundances higher than $+0.20$ dex, and if the 
giants follow the same metallicity distribution, we would expect $\sim 3$ to 
have [Fe/H] abundances higher than $+0.20$ dex (see also Sadakane et al. 2005),
which is clearly not seen.  If the dwarf and giant samples are drawn from the
same parent population, there is only a $0.6\%$ probability of obtaining the
observed metallicity distributions according to a two-sided Kolmogorov-Smirnov
test.


While the sample of giant stars with planetary companions is not statistically 
significant, the indication of a metallicity distribution that possibly 
differs from that of dwarfs is intriguing.  One possible explanation for the 
difference is that the planet-metallicity correlation found for the dwarfs is a 
result of accretion, and the giants have sufficiently diluted their atmospheres 
via evolution-induced mixing.  This conclusion, however, would be difficult to 
reconcile with current results for subgiants (e.g., Fischer \& Valenti 2005).  
Alternatively, planet formation may be sensitive to the mass of the host star, 
independent of metallicity.  The progenitors of G and K giants are early-type MS 
dwarfs and would be more massive than the F, G, and K dwarfs known to have planets. 
Thus, the current results for giants with planets would support a mass-dependent 
planetary formation scenario.  Furthermore, Fischer \& Valenti showed that the 
occurrence of planets around stars increases linearly with stellar mass.  However,
they also found a correlation of rising metallicity with increasing planetary host
mass and argued that the increased occurrence of planets around more massive stars is
most likely spurious. A larger number planets orbiting subgiant and giant stars 
need to be analyzed in order to determine if either of these explanations are 
plausible, or if the difference in metallicity distributions is due to other 
effects, such as different planetary formation mechanisms (e.g., Chauvin et al. 
2005).  Including more evolved stars in RV surveys is highly encouraged.

Our analysis provides the opportunity to place further constraints on the mass of 
HD 13189 from Paper I.  The tracks of \citet{2000A&AS..141..371G} are plotted 
in the \teff-$\log g$ plane in Figure 2; the position of HD 13189 is marked by 
the open star.  From our derived stellar parameters, we find a mass of 
$M_{\star} = 3.5 \; M_{\sun}$ for HD 13189.  Unfortunately, the range of 
possible masses, $M_{\star} =$ 2-6 $M_{\sun}$, is not well-constrained due to 
the uncertainty in the $\log g$ value and is not a significant improvement over 
that given by Paper I.  Adopting a stellar mass of 3.5 $M_{\sun}$, we find a 
minimum companion mass of $14.0 \pm 0.8 \; M_J$, placing the companion at the 
planet/brown dwarf boundary.  HD 13189 and its companion are a potentially 
unique system among planetary systems known today.  The star has one of the 
lowest metallicities and possibly the highest mass of all planetary hosts, and 
the companion likely lies on the planet/brown dwarf boundary and has the largest
semi-major axis for a planet orbiting a significantly metal-poor star.  Further 
investigation of the HD 13189 system is needed to place more stringent 
constraints on the mass of the two objects and should provide unparalleled data 
in the effort to understand the formation and evolution of planetary systems.


\acknowledgments
REU students J.H.K. and M.T. gratefully acknowledge the support of summer research 
through the NSF-REU Award 0353849 to Clemson University.  S.C.S. and J.R.K 
acknowledge support for this work by grant AST 02-39518 to J.R.K. from NSF, as well 
as a generous grant from the Charles Curry Foundation to Clemson University.

\clearpage

\clearpage
\begin{deluxetable}{lcrcrcrcrrcrr}
\tablecolumns{13}
\tablewidth{0pt}
\tablenum{1}
\tablecaption{Equivalent Widths AND Abundances}
\tablehead{
     \colhead{}&
     \colhead{}&
     \colhead{$\lambda$}&
     \colhead{}&
     \colhead{$\chi$}&
     \colhead{}&
     \colhead{}&
     \colhead{}&
     \colhead{$\mathrm{EW}_{\sun}$}&
     \colhead{}&
     \colhead{}&
     \colhead{EW}&
     \colhead{}\\
     \colhead{Species}&
     \colhead{}&
     \colhead{({\AA})}&
     \colhead{}&
     \colhead{(eV)}&
     \colhead{}&
     \colhead{$\log gf$}&
     \colhead{}&
     \colhead{(m{\AA})}&
     \colhead{$\log N_{\sun}$}&
     \colhead{}&
     \colhead{(m{\AA})}&
     \colhead{$\log N$}
     }

\startdata
\ion{Fe}{1}\dotfill && 5633.95 && 4.99 && -0.27 &&  82.4 & 7.59 &&  95.1 & 6.97\\
                    && 5705.47 && 4.30 && -1.57 &&  40.1 & 7.60 &&  82.7 & 7.14\\
                    && 5720.90 && 4.55 && -1.95 &&  20.1 & 7.77 &&  46.4 & 7.30\\
                    && 5731.77 && 4.26 && -1.19 &&  59.7 & 7.54 &&  97.1 & 6.92\\
                    && 5732.30 && 4.99 && -1.50 &&  16.9 & 7.63 &&  31.7 & 7.16\\
\enddata
\tablecomments{Table 1 is presented in its entirety in the electronic
edition of the {\it Astrophysical Journal}.  A portion is provided here for
guidance regarding its form and content.}
\end{deluxetable}

\clearpage
\begin{deluxetable}{lcccr}
\tablecolumns{5}
\tablewidth{0pt}
\tablenum{2}
\tablecaption{Final Results}
\tablehead{
     \colhead{Parameter}&
     \colhead{}&
     \colhead{Value}&
     \colhead{}&
     \colhead{$\sigma$}
     }

\startdata
$T_{\mathrm{eff}} \; (\mathrm{K})\dotfill$ && 4180 && 77\\
$\log g \dotfill$       &&  1.07 && 0.19\\
$\xi \; (\mathrm{km} \; \mathrm{s}^{-1})\dotfill$ && 2.12 && 0.22\\
                                &&       &&     \\
$\mathrm{[Fe/H]}\dotfill$ && -0.58 && 0.04\\
$\mathrm{[O/H]}\dotfill$  && -0.22 && 0.08\\
$\mathrm{[Na/H]}\dotfill$ && -0.31 && 0.06\\
$\mathrm{[Mg/H]}\dotfill$ && -0.21 && 0.05\\
$\mathrm{[Al/H]}\dotfill$ && -0.23 && 0.06\\
$\mathrm{[Si/H]}\dotfill$ && -0.36 && 0.07\\
$\mathrm{[Ca/H]}\dotfill$ && -0.49 && 0.07\\
$\mathrm{[Sc/H]}\dotfill$ && -0.52 && 0.07\\
$\mathrm{[Ti/H]}\dotfill$ && -0.42 && 0.07\\
$\mathrm{[V/H]}\dotfill$  && -0.38 && 0.12\\
$\mathrm{[Cr/H]}\dotfill$ && -0.47 && 0.09\\
$\mathrm{[Mn/H]}\dotfill$ && -0.58 && 0.09\\
$\mathrm{[Ni/H]}\dotfill$ && -0.57 && 0.04\\
\enddata
\end{deluxetable}

\clearpage

\begin{deluxetable}{lcccccc}
\tablecolumns{7}
\tablewidth{0pt}
\tablenum{3}
\tablecaption{Abundance Sensitivities}
\tablehead{
     \colhead{}&
     \colhead{}&
     \colhead{$\Delta T_{\mathrm{eff}}$}&
     \colhead{}&
     \colhead{$\Delta \log g$}&
     \colhead{}&
     \colhead{$\Delta \xi$}\\
     \colhead{Species}&
     \colhead{}&
     \colhead{$\pm 150 \; \mathrm{K}$}&
     \colhead{}&
     \colhead{$\pm 0.25 \; \mathrm{dex}$}&
     \colhead{}&
     \colhead{$\pm 0.30 \; \mathrm{km} \; \mathrm{s}^{-1}$}
     }
     
\startdata
\ion{Fe}{1}\dotfill && $^{+0.04}_{+0.02}$ && $\pm 0.05$ && $\mp 0.09$\\
\ion{Fe}{2}\dotfill && $\mp 0.25$ && $\pm 0.15$ && $\mp 0.06$\\
\ion{O}{1}\dotfill  && $\pm 0.01$ && $\pm 0.08$ && $\mp 0.02$\\
\ion{Na}{1}\dotfill && $\pm 0.15$ && $^{+0.01}_{+0.02}$ && $\mp 0.09$\\
\ion{Mg}{1}\dotfill && $\pm 0.03$ && $\pm 0.02$ && $\mp 0.09$\\
\ion{Al}{1}\dotfill && $\pm 0.13$ && $\pm 0.00$ && $\mp 0.07$\\
\ion{Si}{1}\dotfill && $\mp 0.12$ && $\pm 0.07$ && $\mp 0.04$\\
\ion{Ca}{1}\dotfill && $\pm 0.18$ && $0.00$     && $\mp 0.10$\\
\ion{Sc}{2}\dotfill && $\mp 0.04$ && $\pm 0.11$ && $\mp 0.06$\\
\ion{Ti}{1}\dotfill && $\pm 0.23$ && $0.00$     && $\mp 0.09$\\
\ion{V}{1}\dotfill  && $\pm 0.26$ && $\pm 0.02$ && $\mp 0.15$\\
\ion{Cr}{1}\dotfill && $\pm 0.12$ && $\pm 0.01$ && $\mp 0.08$\\
\ion{Mn}{1}\dotfill && $\pm 0.09$ && $\pm 0.03$ && $\mp 0.02$\\
\ion{Ni}{1}\dotfill && $\pm 0.02$ && $\pm 0.07$ && $\mp 0.10$\\
\enddata
\end{deluxetable}

\clearpage

\begin{figure}
\plotone{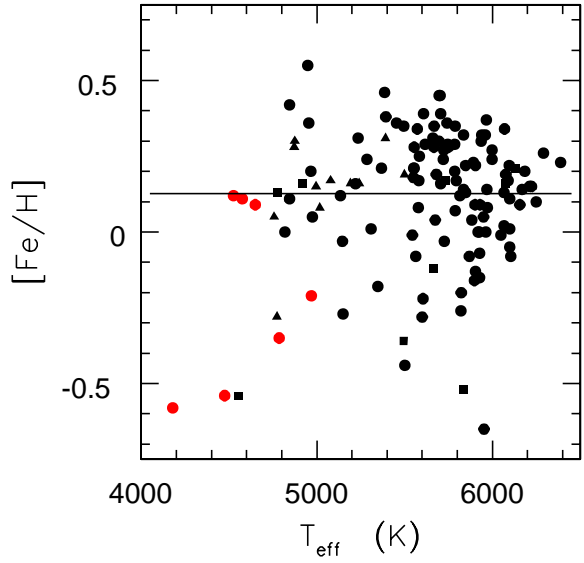}
\caption{Relative Fe abundances of giants (red circles) and dwarfs with 
planetary companions versus \teff.  The sources of the giant abundances and 
\teffs other than those for HD 13189 are given in the text.  Dwarf abundances 
are from Fischer \& Valenti (2005; black circles), Santos, Israelian, \& Mayor 
(2004; black squares), and Santos et al. (2005; black triangles).  The 
horizontal line represents the mean Fe abundance (${\rm [Fe/H]} = +0.13$) of 
the dwarf sample.}
\end{figure}

\clearpage

\begin{figure}
\plotone{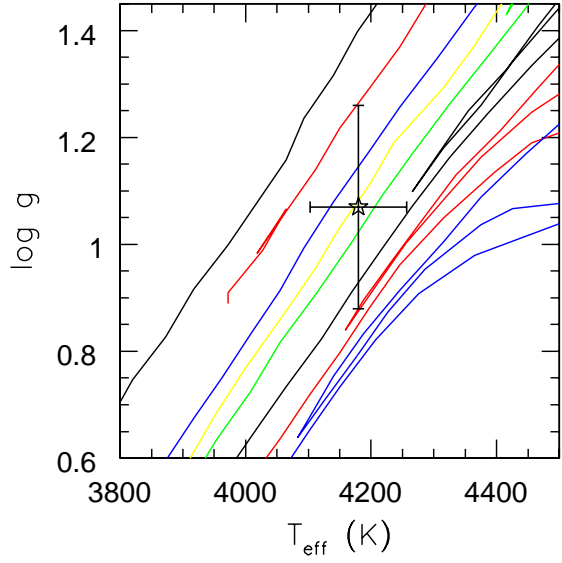}
\caption{Evolutionary tracks of \citet{2000A&AS..141..371G} for (from top left
to bottom right) $M_{\star} = $ 1.5, 2.0, 3.0, 3.5, 4.0, 5.0, 6.0, and 7.0 
$M_{\sun}$.  The tracks are characterized by a metallicity of ${\rm [Fe/H]} = 
-0.50$ or $Z \approx 0.008$.  The open star with error bars marks the location 
of HD 13189 in the \teff-$\log g$ plane.}
\end{figure}


\begin{thebibliography}{}
\bibitem[Allende Prieto et al.(2001)]{2001ApJ...556L..63A} 
Allende Prieto, C., Lambert, D.~L., \& Asplund, M.\ 2001, \apjl, 556, L63
\bibitem[Chauvin et al.(2005)]{2005A&A...438L..29C} Chauvin, G.~ et al.\ 
2005, \aap, 438, L29
\bibitem[Edvardsson et al.(1993)]{1993A&A...275..101E} Edvardsson, B., 
Andersen, J., Gustafsson, B., Lambert, D.~L., Nissen, P.~E., \& Tomkin, J.\ 
1993, \aap, 275, 101
\bibitem[Endl et al.(2003)]{2003AJ....126.3099E} Endl, M., Cochran, W.~D., 
Tull, R.~G., \& MacQueen, P.~J.\ 2003, \aj, 126, 3099
\bibitem[Fischer \& Valenti(2005)]{2005ApJ...622.1102F} Fischer, D.~A. \& 
Valenti, J.\ 2005, \apj, 622, 1102
\bibitem[Fitzpatrick \& Sneden(1987)]{1987BAAS...19.1129F} Fitzpatrick, 
M.~J.~\& Sneden, C.\ 1987, \baas, 19, 1129 
\bibitem[Fulbright(2002)]{2002AJ....123..404F} Fulbright, J.~P.\ 2002, \aj, 
123, 404
\bibitem[Girardi et al.(1996)]{1996A&AS..117..113G} Girardi, L., Bressan, 
A., Chiosi, C., Bertelli, G., \& Nasi, E.\ 1996, \aaps, 117, 113
\bibitem[Girardi et al.(2000)]{2000A&AS..141..371G} Girardi, L., Bressan, 
A., Bertelli, G., \& Chiosi, C.\ 2000, \aaps, 141, 371
\bibitem[Gonzalez(1997)]{1997MNRAS.285..403G} Gonzalez, G.\ 1997, \mnras, 
285, 403
\bibitem[Hatzes et al.(2003)]{2003toed.conf..441H} Hatzes, A.~P., Guenther, 
E., K{\" u}rster, M., \& McArthur, B.\ 2003, ESA SP-539: Earths: DARWIN/TPF 
and the Search for Extrasolar Terrestrial Planets, 441
\bibitem[Hatzes et al.(2005)]{2005A&A...437..743H} Hatzes, A.~P., Guenther, 
E.~W., Endl, M., Cochran, W.~D., D{\"o}llinger, M.~P., \& Bedalov, A.\ 
2005, \aap, 437, 743 (Paper I)
\bibitem[Ida \& Lin(2004)]{2004ApJ...616..567I} Ida, S.~ \& Lin, D.~N.~C.\ 
2004, \apj, 616, 567
\bibitem[Israelian et al.(2001)]{2001Natur.411..163I} Israelian, G., 
Santos, N.~C., Mayor, M., \& Rebolo, R.\ 2001, \nat, 411, 163
\bibitem[Israelian et al.(2003)]{2003A&A...405..753I} Israelian, G., 
Santos, N.~C., Mayor, M., \& Rebolo, R.\ 2003, \aap, 405, 753
\bibitem[Kupka et al.(1999)]{1999A&AS..138..119K} Kupka, F., Piskunov, N., 
Ryabchikova, T.~A., Stempels, H.~C., \& Weiss, W.~W.\ 1999, \aaps, 138, 119 
\bibitem[Montalb{\' a}n \& Rebolo(2002)]{2002A&A...386.1039M} Montalb{\' 
a}n, J.~ \& Rebolo, R.\ 2002, \aap, 386, 1039
\bibitem[Paulson et al.(2004)]{2004AJ....127.3579P} Paulson, D.~B., 
Cochran, W.~D., \& Hatzes, A.~P.\ 2004, \aj, 127, 3579
\bibitem[Pinsonneault et al.(2001)]{2001ApJ...556L..59P} Pinsonneault, 
M.~H., DePoy, D.~L., \& Coffee, M.\ 2001, \apjl, 556, L59
\bibitem[Piskunov et al.(1995)]{1995A&AS..112..525P} Piskunov, N.~E., 
Kupka, F., Ryabchikova, T.~A., Weiss, W.~W., \& Jeffery, C.~S.\ 1995, 
\aaps, 112, 525 
\bibitem[Proffitt \& Michaud(1989)]{1989ApJ...346..976P} Proffitt, C.~R.~ 
\& Michaud, G.\ 1989, \apj, 346, 976
\bibitem[Reddy et al.(2002)]{2002MNRAS.335.1005R} Reddy, B.~E., Lambert, 
D.~L., Laws, C., Gonzalez, G., \& Covey, K.\ 2002, \mnras, 335, 1005
\bibitem[Rice \& Armitage(2005)]{2005ApJ...630.1107R} Rice, W.~K.~M. \& 
Armitage, P.~J.\ 2005, \apj, 630, 1107
\bibitem[Ryabchikova, T.A. et al.(1999)]{rya}Ryabchikova, T.A., Piskunov, N.E.,
Stempels, H.C., Kupka, F., \& Weiss, W.W. 1999, Proc. of the 6th Intern. Colloq.
on Atomic Spectra and Oscillator Strengths, Victoria, BC, Canada, 1998, Physica
Scripta, T83, 162
\bibitem[Sadakane et al.(2002)]{2002PASJ...54..911S} Sadakane, K., Ohkubo, 
M., Takeda, Y., Sato, B., Kambe, E., \& Aoki, W.\ 2002, \pasj, 54, 911 
\bibitem[Sadakane et al.(2005)]{2005PASJ...57..127S} Sadakane, K., Ohnishi, 
T., Ohkubo, M., \& Takeda, Y.\ 2005, \pasj, 57, 127
\bibitem[Santos et al.(2003)]{2003A&A...398..363S} Santos, N.~C., 
Israelian, G., Mayor, M., Rebolo, R., \& Udry, S.\ 2003, \aap, 398, 363
\bibitem[Santos et al.(2004)]{2004A&A...415.1153S} Santos, N.~C., 
Israelian, G., \& Mayor, M.\ 2004, \aap, 415, 1153
\bibitem[Santos et al.(2005)]{2005A&A...437.1127S} Santos, N.~C., 
Israelian, G., Mayor, M., Bento, J.~P., Almeida, P.~C., Sousa, S.~G., \& 
Ecuvillon, A.\ 2005, \aap, 437, 1127
\bibitem[Sato et al.(2003)]{2003ApJ...597L.157S} Sato, B., et al.\ 2003, 
\apjl, 597, L157
\bibitem[Schuler et al.(2005a)]{forb}Schuler, S.~C., Hatzes, A.~P., King, J.~R.,
K{\"u}rster, M., Booesgaard, A.~M., and The, L.~-S. \ 2005a, AJ, submitted
\bibitem[Schuler et al.(2005b)]{trip}Schuler, S.~C., King, J.~R., Terndrup, 
D.~M., Pinsonneault, M.~H., Murray, N., \& Hobbs, L.~M. \ 2005b, \apj, submitted
\bibitem[Setiawan et al.(2003)]{2003A&A...397.1151S} Setiawan, J., 
Pasquini, L., da Silva, L., von der L{\" u}he, O., \& Hatzes, A.\ 2003, 
\aap, 397, 1151
\bibitem[Setiawan et al.(2005)]{2005A&A...437L..31S} Setiawan, J.~ et al.\ 
2005, \aap, 437, L31
\bibitem[Smith et al.(2001)]{2001AJ....121.3207S} Smith, V.~V., Cunha, K., 
\& Lazzaro, D.\ 2001, \aj, 121, 3207
\bibitem[Sneden(1973)]{1973ApJ...184..839S} Sneden, C.\ 1973, \apj, 184, 
839
\bibitem[Thevenin(1990)]{1990A&AS...82..179T} Thevenin, F.\ 1990, \aaps, 
82, 179
\end{thebibliography}
\end{document}